\documentclass[%
 aip,
 jmp,%
 amsmath,amssymb,
 reprint,%
]{revtex4-1}

\usepackage{graphicx}
\usepackage{dcolumn}
\usepackage{bm}

\begin{document}


\title{Two-shot measurement of spatial coherence}

\author{Abhinandan Bhattacharjee, Shaurya Aarav, and Anand K. Jha}
\email{akjha9@gmail.com}
 
\affiliation{Department of Physics, Indian Institute of Technology Kanpur, Kanpur, UP 208016, India 
}%


\date{\today}

\begin{abstract}
We propose and demonstrate an interferometric scheme for measuring the two-dimensional two-point cross-spectral density function in a two-shot manner. Our scheme comprises a Michelson interferometer with a converging lens in one of the arms of the interferometer, and the cross-spectral density function of an input optical field gets encoded in the intensity distribution of the output interferograms. This scheme works for any cross-spectral density function that is real and that depends on the spatial coordinates only through their difference. Using this scheme, we report measurements of several lab-synthesized cross-spectral density functions with very good agreement with theory. Our measurement technique can be very important for applications that are based on utilizing the partial spatial coherence properties of optical fields.

\end{abstract}

\maketitle

Spatial coherence referes to the correlation between a pair of space points in an optical field. It is quantified through the so-called cross-spectral density function. Fields having partial spatial coherence offer a wide range of applications including wide-field optical coherence tomography (OCT) \cite{karamata2004optlett,dhalla2010optlet}, imaging
through turbulence \cite{redding2012natphot,redding2015pnas}, coherence holography \cite{naik2009optexp}, photon correlation holography \cite{naik2011optexp}, optical communication \cite{ricklin2002josaa} and particle trapping \cite{zhao2009optexp,dong2012pra}. For all of these applications, a fast and accurate way of measuring the cross-spectral density function is an essential requirement.

There are several different ways of measuring the cross-spectral density function of an optical field. The Young's double-slit interferometer \cite{mandel&wolf1995,zernike1938physica,turunen1991josaa} and its variants \cite{santarseiro2006ol} are among the most commonly used techniques. However, the techniques based on Young's double-slit interferometry have several drawbacks. First of all, in order to measure the cross-spectal density function with increased resolution, one requires progressively narrower slits. This requirement makes such techniques very difficult to use for light fields with very low intensities or to generalize them for measuring two-dimensional functions. Furthermore, the measurement of cross-spectral density functions using such techniques requires multiple measurements with varying slit separations. This increases the measurement time as well as the stability requirements for the interferometers. Other schemes for measuring the cross-spectral density function include shearing interferometry \cite{efimov2013optlet,iaconis1996optlet}, phase-space tomography \cite{nugent1992prl, smithey1993prl}, the schemes based on free space propagation \cite{rydberg2007optexp, petruccelli2013optexp} and the schemes based on scanning a small obstacle over the test plane
and then measuring the resulting radiant intensity.\cite{wood2014optlet, sharma2016optexp}. However, these methods are either not suitable for low-intensity fields or require multiple measurements and are thus unsuitable for measuring two-dimensional functions in an efficient manner. A scheme proposed by Wessely {\it et al.} \cite{wessely1970josa} does measure the two-dimensional cross-spectral density function in a single shot manner without requiring multiple measurements; however, due to the finite edge-width of the prisms used in the scheme, the scheme misses out some information and as a result does not measure the entire cross-spectral density function.  

In contrast, in this letter, we propose and demonstrate an image-inversion based interferometric technique for measuring the two-dimensional cross-spectral density functions in a two-shot manner. Our technique is the spatial analog of the technique recently proposed and implemented \cite{kulkarni2017natcomm} in the orbital angular momentum basis for measuring the angular coherence function \cite{jha2011pra}, and it works for any two-dimensional cross-spectral density function that is real and that depends on the spatial coordinates only through their difference.

\begin{figure*}[t!]
\includegraphics{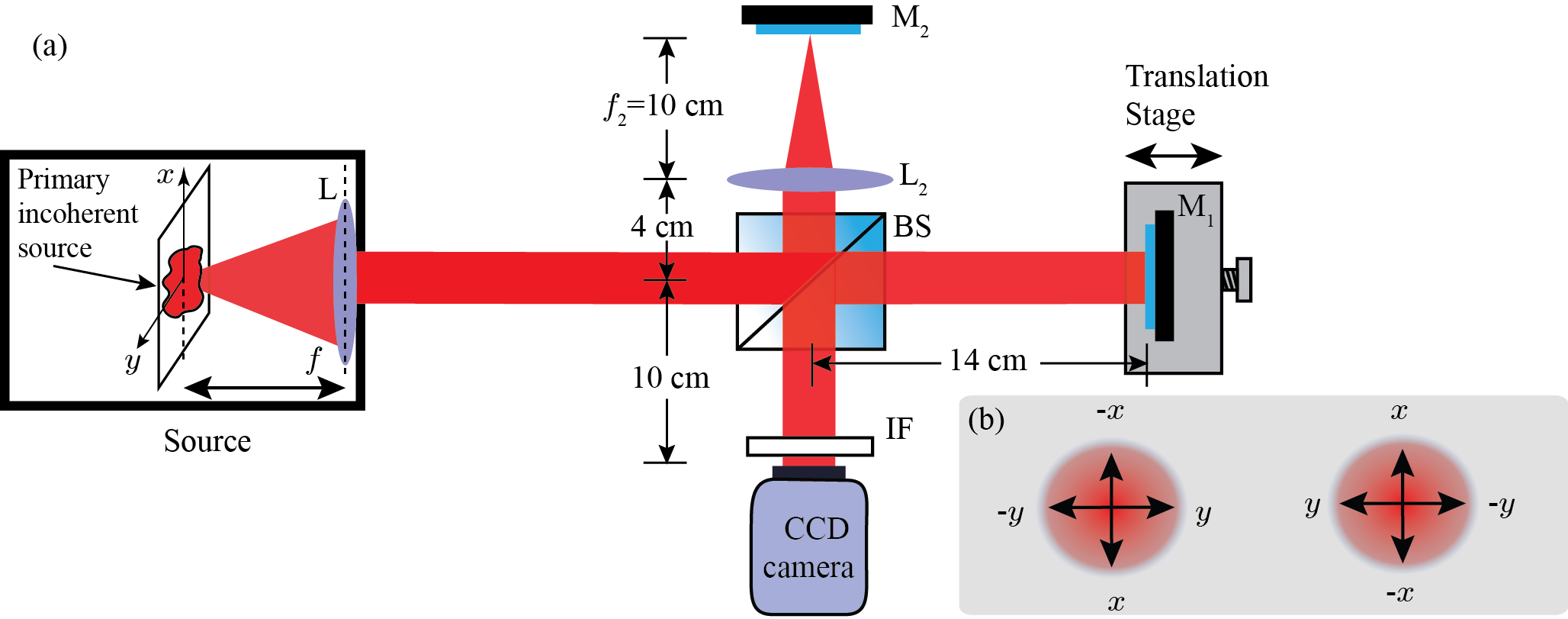}
\centering
\caption{(color online) (a) Schematic diagram of the experimental
setup. The primary incoherent source is kept at the back focal plane of a converging lens L with focal length $f=200$ cm. The mirror ${\rm M_2}$ is kept at the back focal plane of the converging lens ${\rm L_2}$ of focal length $f_2=10$ cm. The length of each interferometric arm is about 14 cm and the CCD camera is kept at about 10 cm from the beam splitter (BS). An interference filter (IF) centered at 632.8 nm having a wavelength-bandwidth of 10 nm is used before the CCD camera. The spatially partially coherent field exiting the lens L ends up having the cross-spectral density function that depends on the spatial coordinates only through their difference \cite{aarav2017pra}. (b) The two interfering wavefronts at the CCD camera plane. The wavefront coming through the interferomtric arm having lens  ${\rm L_2}$  is inverted in both $x$ and $y$ directions compared to the wavefront coming through the arm having no lens. In the above figure, we have used the following abbreviations:  BS stands for beam splitter, M for mirror, L for converging lens, and IF for interference filter. 
 }\label{fig1}
\end{figure*}

Figure \ref{fig1} illustrates our proposed method and shows the schematic diagram of our experimental setup. The source generates a  spatially partially coherent field. We represent the field produced by the source in any given realization by $E_{\rm in}({\bm\rho})$. The cross-spectral density function of the field, which quantifies the spatial coherence in the field at the two space points ${\bm \rho_1}$ and ${\bm \rho_2}$, is defined as $W({\bm{\rho}_1}, {\bm{\rho}_2})=\langle E^*_{\rm in}({\bm\rho_1})E_{\rm in}({\bm \rho_2})\rangle$,  where $\langle\cdots\rangle$ denotes the ensemble average over many different realizations of the field. We aim to measure the cross-spectral density function using the interferometer shown in Fig.~\ref{fig1}(a). The interferometer has two arms. One arm contains a mirror while the other arm contains a converging lens along with a mirror kept at the back focal plane of the converging lens. For a collimated field, the lens produces an inverted wavefront at the mirror which is reflected back onto the lens. After reflection, the inverted wavefront is collimated back by the lens producing a wavefront which is inverted in both $x$ and $y$ directions with respect to the incoming collimated field, that is, $\bm\rho\rightarrow-\bm\rho$. The two interfering wavefronts at the detection plane (CCD camera) has been illustrated in Fig.~\ref{fig1}(b). The field at the output port of the interferometer can therefore be written as
\begin{align}
E_{\rm out}({\bm{\rho}})=&\sqrt{k_1}E_{in}(\bm\rho)e^{i(\omega_0t_1+\beta_1)} \notag \\
& \qquad+\sqrt{k_2}E_{in}(-\bm\rho)e^{i(\omega_0t_2+\beta_2)}.
\end{align}
Here, $t_1$ and $t_2$ denote the times taken by the field to travel through the two arms of the interferometer; $\omega_0$ is the frequency of the field; $\beta_1$ and $\beta_2$ are the phases other than the dynamical phases acquired in both the arms; $k_1$ and $k_2$ are the scaling constants in the two arms. The intensity $I_{\rm out}(\bm\rho)$ at the output port of the interferometer is given by $I_{\rm out}(\bm\rho)=\langle E^*_{\rm out}(\bm\rho)E_{\rm out}(\bm\rho)\rangle$ and can be shown to be
\begin{align}
I_{\rm out}({\bm{\rho}})=& k_1 \langle E^*_{\rm in}(\bm\rho)E_{\rm in}(\bm\rho)\rangle+k_2 \langle E^*_{\rm in}(-\bm\rho)E_{\rm in}(-\bm\rho)\rangle \notag\\ &+\sqrt{k_1k_2}\langle E^*_{\rm in}(\bm\rho)E_{\rm in}(-\bm\rho)\rangle e^{i\delta} + {\rm c.c.} \label{I_out},
\end{align}
where $\delta=\omega_0(t_2-t_1)+(\beta_2-\beta_1)$. We assume that the cross-spectral density function  $\langle E^*_{\rm in}(\bm\rho)E_{\rm in}(-\bm\rho)\rangle=W(\bm\rho,-\bm\rho)$ produced by our source depends on the spatial coordinates only through their difference ${\bm \Delta\rho}={\bm \rho_1}-{\bm \rho_2}$.  As a result, we write $W(\bm\rho,-\bm\rho)$ as $W(2\bm\rho)$. We also write $\langle E^*_{\rm in}(\bm\rho)E_{\rm in}(\bm\rho)\rangle=I(\bm\rho)$, and $\langle E^*_{\rm in}(-\bm\rho)E_{\rm in}(-\bm\rho)\rangle =I(-\bm\rho)$. Therefore $I_{\rm out}({\bm{\rho}})$ can be written as
\begin{align}
I_{\rm out}&({\bm{\rho}})=k_1 I(\bm\rho)+ k_2 I(-\bm\rho) \notag \\ + &2\sqrt{k_1k_2} \{{\rm Re}[W(2\bm\rho)]\cos\delta -{\rm Im}[W(2\bm\rho)]\sin\delta\} \label{I_out}.
\end{align}
Here ${\rm Re}[W(2\bm\rho)]$ and ${\rm Im}[W(2\bm\rho)]$ denote the real and imaginary parts of the cross-spectral density function, respectively. Also, since the cross spectral density function depends on ${\bm \Delta\rho}$ only, we have $I(\bm\rho)=I(-\bm\rho)=W(\bm\rho, \bm\rho)=C$, where $C$ is a constant. It is clear from the above equation that the output intensity $I_{\rm out}(\bm\rho)$ has the cross-spectral density function $W(2\bm\rho)$ encoded in it. If the cross-spectral density function is real and if we know the values of $k_1$, $k_2$, $I(\bm\rho)$ and $\delta$ then in principle a single-shot measurement of the output interferogram $I_{\rm out}(\bm\rho)$ will yield the cross-spectral density function $W(2\bm\rho)$ of the field. However, it is in general very difficult to obtain $W(2\bm\rho)$ this way because of the requirement that $k_1$, $k_2$, $I(\bm\rho)$ and $\delta$ should be known precisely. Any error in the knowledge of these quantities introduces error in the estimation of the cross-spectral density function. Furthermore, there are wavefront errors introduced by the interferometer which also degrade the fidelity of the estimation. Nevertheless, it has been shown in Ref.~\cite{kulkarni2017natcomm} that if, instead of one, two suitable output interferograms are collected then not only the estimation becomes independent of wavefront errors but also there remains no need to know $k_1$, $k_2$, $I(\bm\rho)$ and $\delta$. This can be illustrated as follows. Suppose the experimentally measured output intensity $\bar{I}^{\delta}_{\rm out}(\bm\rho)$ at $\delta$ contains some background $I^{\delta}_b(\bm\rho)$ in addition to the signal $I_{\rm out}(\bm\rho)$. Therefore, $\bar{I}^{\delta}_{\rm out}(\bm\rho)$ can be written as
\begin{figure*}[t!]
\centering
\includegraphics{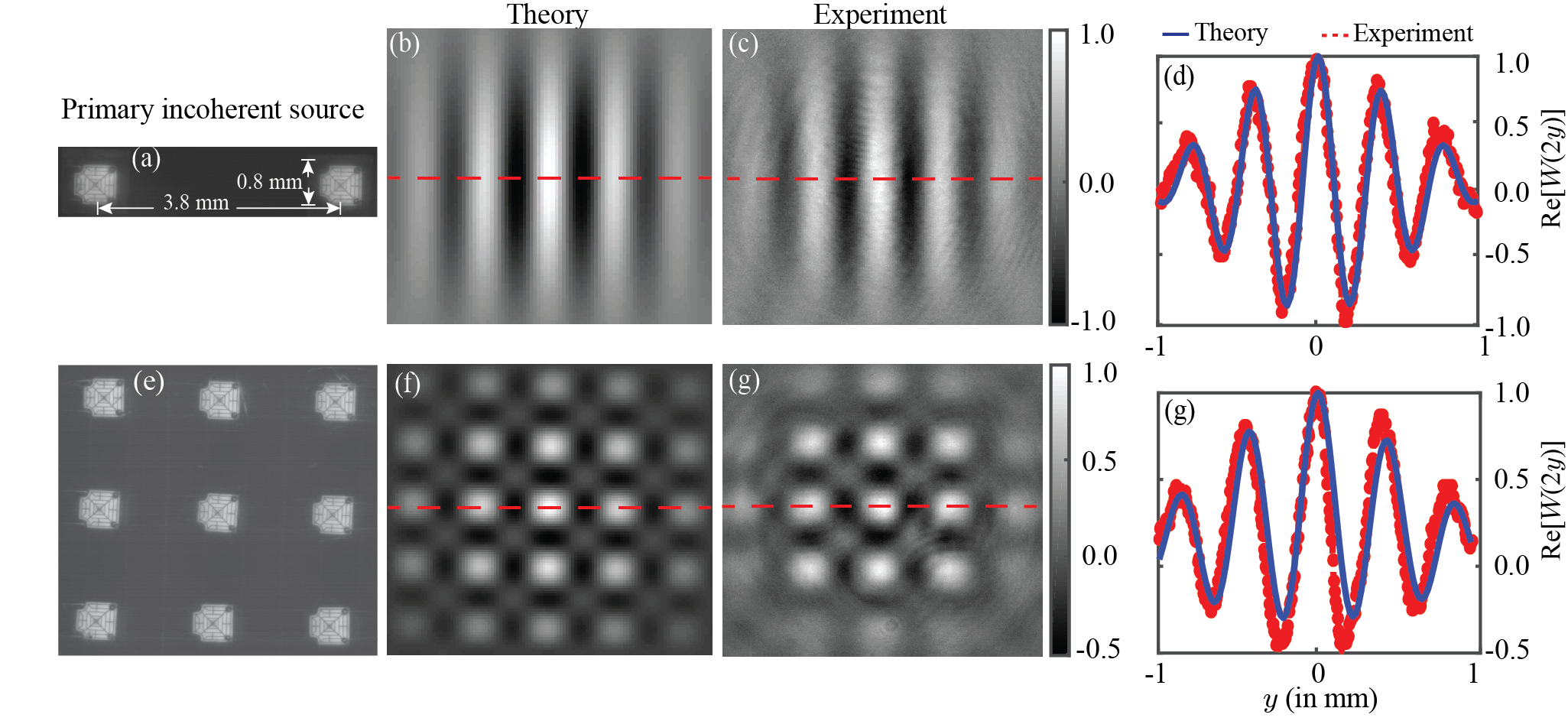}
\caption{(color online) (a) and (e) CCD camera images of two separate primary incoherent sources. (b) and (f) The theoretical cross-spectral density function ${\rm Re}[W(2\bm\rho)]$ of the spatially partially coherent fields produced by the combination of the primary incoherent source and the converging lens. (c) and (g) The experimentally measured ${\rm Re}[W(2\bm\rho)]$. (d) and (h) Plots of the one-dimensional cuts along the $y$-direction of the theoretical and experimental cross-spectral density functions. The theoretical and experimental plots have been scaled such that the maximum of ${\rm Re}[W(2\bm\rho)]$ is one. 
 }\label{fig2}
\end{figure*}
\begin{align}
\bar{I}^{\delta}_{\rm out}&({\bm{\rho}})=I^{\delta}_b(\bm\rho)+k_1 C+ k_2 C \notag \\ + &2\sqrt{k_1k_2} \{{\rm Re}[W(2\bm\rho)]\cos\delta -{\rm Im}[W(2\bm\rho)]\sin\delta\}.
\end{align}
Now, let us assume that we have two output interferograms with intensities $\bar{I}^{\delta_c}_{\rm out}(\bm\rho)$ and 
$\bar{I}^{\delta_d}_{\rm out}(\bm\rho)$ measured at $\delta=\delta_c$ and $\delta=\delta_d$, respectively. The difference $\Delta\bar{I}_{\rm out}(\bm\rho)=\bar{I}^{\delta_c}_{\rm out}(\bm\rho)-\bar{I}^{\delta_d}_{\rm out}(\bm\rho)$ in the intensities of the two interferograms is therefore given by
\begin{align}
\Delta\bar{I}_{\rm out}(\bm\rho)=&\Delta{I}_b(\bm\rho) +2\sqrt{k_1k_2} \notag\\& \times \{ {\rm Re}[W(2\bm\rho)](\cos\delta_c-\cos\delta_d) \notag\\& \qquad\qquad - {\rm Im}[W(2\bm\rho)](\sin\delta_c-\sin\delta_d) \},
\end{align}
where $\Delta{I}_b(\bm\rho)=I^{\delta_c}_b(\bm\rho)-I^{\delta_d}_b(\bm\rho)$ is the difference in background intensities. We assume that the background does not vary from shot to shot, that is, $\Delta{I}_b(\bm\rho)\approx 0$. Furthermore, we assume that the cross-spectral density function is either completely real or has a negligible imaginary part. Now, along with these assumptions, if we measure the two interferograms at $\delta_c\approx 0$ and $\delta_d \approx \pi$, we have ${\rm Im}[W(2\bm\rho)](\sin\delta_c-\sin\delta_d) \ll {\rm Re}[W(2\bm\rho)](\cos\delta_c-\cos\delta_d)$, and thus $\Delta\bar{I}_{\rm out}(\bm\rho)$ becomes effectively proportional to the real part of the cross-spectral density function, that is, 
\begin{align}
\Delta\bar{I}_{\rm out}(\bm\rho)\propto {\rm Re}[W(2\bm\rho)].
\end{align}
Therefore, by measuring the difference intensity $\Delta\bar{I}_{\rm out}(\bm\rho)$, one can directly measure the real part of the cross-spectral density function of the input field. We note that if the intensity of the field in Eq.~(\ref{I_out}) is a constant, its cross-spectral density function can be measured using our method for any pair of space points in the field. However, in situations in which the intensity is not a constant but $I(\bm\rho)=I(-\bm\rho)$, our method can measure the cross-spectral density function around $\bm\rho=0$.

We further note that the above formalism has been worked out for a cross-spectral density function that is either completely real or that has a negligible imaginary part. A cross-spectral density function can in general be complex. For such cross-spectral density functions one can work out a two-shot formalism that is analogous to the one presented in the methods section of Ref. \cite{kulkarni2017natcomm}. However, in contrast to the above formalism, the analogous formalism would require $\delta_c$ and $\delta_d$ to be known precisely.

We now report our experimental measurements of spatially partially coherent fields using the proposed scheme. As discussed above, our scheme works for cross-spectral density functions that depend on the spatial coordinates only through their difference ${\Delta\bm \rho}$. There are several methods for producing such fields \cite{takeda2005optexp, turunen1991josaa}. A very efficient way of generating such fields have been reported very recently \cite{aarav2017pra}, in which a spatially incoherent primary source is placed at the back focal plane of a converging lens (see Fig.~\ref{fig1}) and as a consequence the field exiting the lens ends up having the cross-spectral density function given by \cite{aarav2017pra}.
\begin{align}
W({\bm{\rho}_1}, {\bm{\rho}_2})\rightarrow W(\Delta\bm\rho)=\int_{-\infty}^{\infty} I(\bm{q}) e^{-i\bm{q.\Delta\rho}}d\bm{q}, \label{csd_ss}
\end{align} 
where $I(\bm q)$ is the spectral density of the field exiting the lens and is proportional to the intensity $I_s(\bm\rho'_s)$ of the primary incoherent source \cite{aarav2017pra}, where $\bm\rho'_s$ represent the spatial coordinates at the plane of the primary incoherent source while $\bm\rho$ represent the spatial coordinates at a plane after the converging lens. The cross-spectral density function $W(\Delta\bm\rho)$ depends only on ${\bm\Delta\rho}={\bm\rho_1}-{\bm\rho_2}$ and is the Fourier transform of $I(\bm q)$. Thus it is proportional to the Fourier transform of the source intensity $I_s(\bm \rho'_s)$. We note that the cross-spectral density function of Eq.~(\ref{csd_ss}) represents a field that is both spatially-stationary and propagation invariant \cite{aarav2017pra}. We further note that when $I(\bm q)$ is a symmetric function, $W(\Delta\bm\rho)$ is real. For any real source $I(\bm q)$ cannot entirely be symmetric. However, we assume that the spectral density $I(\bm q)$ of our source is almost symmetric such that $W(\Delta\bm\rho)$ has a negligible imaginary part.

In our experiments, we use a commercially available $9$-W planar light emitting diode (LED) bulb as the primary incoherent source. The LED bulb consists of 9 separate LEDs arranged in a 3$\times$3 grid (see Fig.~\ref{fig2}(e)). The primary source in Fig.~\ref{fig2}(a) is obtained by covering the remain 7 LEDs. The individual LEDs are of dimensions $0.8\times 0.8$ mm and the separation between two nearest LEDs is $1.9$ mm. The source is kept at the back focal plane of lens L having focal length $f=200$ cm. The mirror ${\rm M_2}$ is kept at the back focal plane of the converging lens ${\rm L_2}$ of focal length $f_2=10$ cm. The length of each interferometric arm is about 14 cm and the CCD camera is kept at about 10 cm from the beam splitter (BS). An interference filter (IF) centered at 632.8 nm having a wavelength-bandwidth of 10 nm is used before CCD camera. Figure \ref{fig2} shows our experimental results. Figures \ref{fig2}(a) and \ref{fig2}(e) are the CCD camera images of the two separate primary incoherent sources used. Figures \ref{fig2}(b) and \ref{fig2}(f) are the theoretical cross-spectral density functions of the spatially partially coherent field generated by the combination of the primary incoherent source and the converging lens. These theoretical plots have been generated by first performing the Fourier transform of  Eq.~(\ref{csd_ss}) with intensity $I(\bm\rho')$ of the images in  Figs \ref{fig2}(a) and \ref{fig2}(e) and then taking the real parts. Figures.~\ref{fig2}(c) and \ref{fig2}(g) show the experimentally measured ${\rm Re}[W(2\bm\rho)]$ through our two-shot technique, by collecting suitable interferograms at two different values of $\delta$, in each case. In our experiment, $\delta$ was varied by manually moving the translation stage, and the sets of two interferogram images were collected with $\delta_c \approx 0$ and $\delta_d\approx \pi$. In order to compare our experimental results with theory, we plot in figs.~\ref{fig2}(d) and \ref{fig2}(h) the one-dimensional cuts along $y$-direction of the theoretical and experimental cross-spectral density functions. The theoretical and experimental plots have been scaled such that the maximum of ${\rm Re}[W(2\bm\rho)]$ is one. We find very good agreement between the theory and experiment. This also verifies our assumption that the spectral density $I(\bm q)$ produced by our source is almost symmetric and thus the imaginary part of the cross-spectral density function is negligible. The slight mismatch between the theory and experiment can be attributed to the very low but finite shot-to-shot background variations and to the negligible but finite imaginary part of the cross-spectral density function. We believe that the finite shot-to-shot background can be minimized even further if the phase difference $\delta$ is varied in an automated manner.

In summary, in this letter, we have proposed and demonstrated a scheme for measuring the two-dimensional two-point cross-spectral density function of optical fields in a two-shot manner. We have reported the measurements of a few lab-synthesized cross-spectral density functions with very good agreement with theory. Our measurement technique overcomes the limitations of the conventional interferometers for measuring the cross-spectral density function in that it yields the entire cross-spectral density function using just two shots, is insensitive to background noise, and does not require precise knowledge of experimental parameters. We expect our technique to have important implications for applications such as correlation holography and wide-field OCT that are based on utilizing the partial spatial coherence properties of optical fields.

We thank Girish Kulkarni for fruitful discussions and acknowledge financial support through an initiation grant no. IITK /PHY /20130008 from Indian Institute of Technology (IIT) Kanpur, India and through the research grant no. EMR/2015/001931 from the Science and Engineering Research Board (SERB), Department of Science and Technology, Government of India.


\end{document}